# Giant magnetocaloric effect in isostructural MnNiGe-CoNiGe system by establishing a Curie-temperature window

E. K. Liu[a], H. G. Zhang[a], G. Z. Xu[a], X. M. Zhang[a], R. S. Ma[a], W. H. Wang[a],

J. L. Chen[a], H. W. Zhang[a], G. H. Wu[a,1], L. Feng[b], X. X. Zhang[c]

[a] State Key Laboratory for Magnetism, Beijing National Laboratory for Condensed Matter Physics, Institute of Physics, Chinese Academy of Sciences, Beijing 100190, China

[b] College of Physics and Optoelectronics, Taiyuan University of Technology, Taiyuan 030024, China

[c] Core Labs, King Abdullah University of Science and Technology (KAUST), Thuwal 23955-6900, Saudi Arabia

**Abstract**: An effective scheme of isostructural alloying was applied to establish a Curie-temperature window in isostructural MnNiGe-CoNiGe system. With the simultaneous accomplishment of decreasing structural-transition temperature and converting antiferromagnetic martensite to ferromagnetic state, a 200 K Curie-temperature window was established between Curie temperatures of austenite and martensite phases. In the window, a first-order magnetostructural transition between paramagnetic austenite and ferromagnetic martensite occurs with a sharp jump in magnetization, showing a magnetic entropy change as large as -40 J kg$^{-1}$ K$^{-1}$ in a 50 kOe field change. This giant magnetocaloric effect enables Mn$_{1-x}$Co$_x$NiGe to become a potential magnetic refrigerant.

---

[1] Author to whom correspondence should be addressed; E-mail: ghwu@iphy.ac.cn

Solid-state magnetic refrigeration has increasingly become attractive as a cooling technology due to its many advantages.[1,2] Among diverse magnetic-cooling materials, some undergo first-order magnetostructural transitions with remarkable changes in their structural and magnetic symmetries, which result in considerable magnetocaloric effects (MCEs).[3-9] One such magnetostructural transition is the ferromagnetic martensitic transformation (FMMT), which has been found in cubic Heusler alloys[10] and hexagonal *MM'X* (*M*, *M'* = transition metals, *X* = Si, Ge, Sn) compounds.[11] In FMMT, the martensitic structural transformation is coupled with a magnetic-state transition, exhibiting magnetostructural coupling. It is well accepted that this coupling and the large magnetization difference ($\Delta M$) between the austenite and martensite phases play important roles in MCE.[12-15] Tuning these two aspects is thus a central task in developing magnetocaloric materials with FMMTs.

As an *MM'X* compound, MnNiGe crystallizes in a Ni$_2$In-type hexagonal structure (*P*6$_3$/*mmc*, 194).[16] A first-order martensitic structural transformation (MT) occurs at $T_t$=470 K to the TiNiSi-type orthorhombic structure (*Pnma*, 62).[16] The MT occurs in the paramagnetic (PM) state so that there is no appreciable $\Delta M$. At a lower temperature of 346 K (Néel temperature, $T_N^M$), the martensite phase undergoes a magnetic transition from the PM state to a spiral antiferromagnetic (AFM) one.[17,18] The structural and magnetic transitions in this compound cannot couple together, as shown in Fig. 1 (see $T_t$ and $T_N^M$ for MnNiGe). Many efforts have been made in different ways to realize the desired magnetostructural coupling.[19-22] Recently, Curie-temperature windows (TW)[9,23-28] have been found between the Curie temperatures of the austenite ($T_C^A$) and martensite ($T_C^M$) in $MM'X$ alloys that yields a desired magnetostructural coupling as well as a large $\Delta M$. From the view of the TW, we suggest that by introducing Co atoms to MnNiGe the desired results may also be obtained for a potential magnetocaloric material. A previous report in the 1980s[29] verified that substituting Co for Ni in MnNiGe converts the spiral AFM structure to a FM one by Co-Mn exchange interactions (Fig. 1(a), yellow region below $T_N^M/T_C^M$). However, this substitution failed to realize the magnetostructural coupling. In the series of compositions, as shown in Fig. 1(a), $T_t$ remained above 440 K. There was a wide temperature gap of about 100 K between $T_t$ and the magnetic-ordering temperature ($T_N^M/T_C^M$) (Fig. 1(a), blue region). This result may be understood based on the fact that the two end members (MnNiGe and MnCoGe) are $Ni_2In$-type hexagonal isostructures in which both Ni and Co occupy 2d sites,[29,30] and both isostructures have high-temperature MTs (470 K for MnNiGe, 460 K for MnCoGe ).[16] Therefore, substituting Co for Ni at the same $M'$ (2d) sites could not give rise to a remarkable impact to the phase stability to decrease $T_t$ of the MnNiGe-MnCoGe system.

Thus, we applied the same consideration to the MnNiGe-CoNiGe system and found that a promising magnetostructural coupling may emerge. This is because the isostructural CoNiGe compound has a stable $Ni_2In$-type austenite structure without MT.[31,32] As shown in Fig. 1(b), $T_t$ is likely to decrease to below $T_N^M$ (even below $T_C^A$) and the temperature gap (blue region) would then close. Meanwhile, it is also desirable to form a strong FM phase in AFM martensite by introducing Co-Mn exchange interactions. A TW is thus expected to be opened between $T_N^M$ ($T_C^M$) and $T_C^A$ (Fig. 1(b)). In this letter, we alloyed CoNiGe with MnNiGe and found that magnetostructural coupling as well as a large $\Delta M$ can be effectively attained in an established TW. Through this process, we designed and realized a giant MCE in MnNiGe-CoNiGe system.

Polycrystalline ingots of $Mn_{1-x}Co_xNiGe$ ($x$ = 0, 0.1, 0.2 and 0.3) were prepared by arc melting pure metals in a highly pure argon atmosphere. The ingots were annealed at 1123 K in an evacuated quartz tube for five days and then cooled slowly to room temperature. The structures of the materials were identified by powder x-ray diffraction (XRD) analysis with Cu-Kα radiation. The phase transitions of the samples were detected by differential scanning calorimetry (DSC) (for $x$ = 0) and low-field thermomagnetization, M ($T$), curves (for $x$ = 0.1, 0.2, 0.3). Magnetic measurements were performed on a superconducting quantum interference device (SQUID).

Room-temperature XRD patterns of $Mn_{1-x}Co_xNiGe$ ($x$ = 0, 0.1, 0.2 and 0.3) samples are shown in Fig. 2. The diffraction patterns confirm that Co-substituted samples crystallize a $Ni_2In$-type hexagonal structure at room temperature. On the other hand, the sample with $x$ = 0 has a TiNiSi-type orthorhombic structure. These results indicate that $T_t$ decreases below room temperature by alloying CoNiGe with MnNiGe. Compared with the reported MnNiGe-MnCoGe system shown in Fig. 1, the manipulation of the phase stability by isostructural alloying works well in the MnNiGe-CoNiGe system. The lattice parameters of two phases are listed in Table I. When the Co content increases, the lattice parameters of the austenite phase decrease, as shown in the inset to Fig. 2. This decrease can be ascribed to the substitution of small-size Co ($r$ = 1.26 Å) for large-size Mn ($r$ = 1.39 Å).

To identify the structural and magnetic transitions, we measured the low- and high-field M ($T$) curves of Mn$_{1-x}$Co$_x$NiGe ($x$ = 0, 0.1, 0.2 and 0.3) samples as shown in Figs. 3(a) and 3(b). These figures show that $T_t$ decreases with increasing Co content. For $x$ = 0, the transformation occurs at 460 K and the martensite antiferromagnetically orders at $T_N^M$ = 356 K (Table I) (DSC analysis of $T_t$ and $T_N^M$ shown in supplementary Fig. S1).[36] Furthermore, below $T_N^M$, the martensite exhibits moderate magnetization with values from 30 to 45 emu g$^{-1}$ in the high field of 50 kOe due to the instability of spiral AFM magnetic structures, as shown in Fig. 3(b). These results are consistent with those reported for stoichiometric MnNiGe.[18,21,29]

A sharp, first-order MT from PM austenite to FM martensite is observed at 236 K with a small hysteresis of 11 K on the sample with $x$ = 0.1 (Figs. 3(a) and 3(b), Table I), indicating that the desired magnetostructural coupling is established at this Co content. In addition, a $\Delta M$ as large as 53 emu g$^{-1}$ in the 50 kOe field is found during the phase transition (Fig. 3(b), Table I), which suggests the possibility of the desired magnetic-field-induced MCEs. Even this small amount of Co content introduces enough FM coupling against the AFM interaction in the originally AFM-coupled martensite matrix. At the same time, the martensite exhibits very interesting magnetic behavior at low temperatures. Below 100 K, the M ($T$) curve increases with decreasing temperature in a 1 kOe field, whereas it decreases as the temperature decreases when a 50 kOe field is applied (supplementary Fig. S2).[36] This complex behavior indicates a striking competition between FM and AFM exchange interactions due to Co substitution in this sample.

For $x = 0.2$, the MT is observed at about 93 K, which is below the Curie temperature of austenite ($T_C^A$ = 129 K) (Fig. 3(a), Table I). Below the MT, the magnetization of this sample increases since the magnetic moments of martensite are larger than those of austenite. This characteristic found in $MM'X$ alloys[9,11,16] is very different from many Heusler alloys.[13-15] The magnetostructural transition from FM austenite to FM martensite provides only a small $\Delta M$ of 26 emu g$^{-1}$ (Fig. 3(b), Table I), which is half that for $x = 0.1$. Between $T_N^M$ = 356 K of $x = 0$ and $T_C^A$ = 129 K of $x = 0.2$, a TW arises with a width of about 200 K, similar to the reported TWs for MnCoGe$_{1-x}$Sn$_x$,[24] Mn$_{1-x}$CoGe[23] and MnNiGe:Fe.[9] These TWs originate because the value of $T_C^M$ is much higher than that of $T_C^A$ in $MM'X$ alloys.[16,29,30] Within the TW, when the transformation occurs, the parent phase is still in the PM state while the martensite is already in the FM state. Thus, a PM-to-FM MT occurs with a large $\Delta M$. For $x = 0.3$, the magnetostructural transition decouples as the MT vanishes, leaving a magnetic ordering at $T_C^A$ = 124 K (Figs. 3(a) and 3(b), Table I).

To further analyze the ferromagnetism of the martensite phase, we measured the magnetization, M (H), curves of Mn$_{1-x}$Co$_x$NiGe samples ($x$ = 0, 0.1, 0.2) at 5 K, as shown in Fig. 4. The three samples were all in martensite state. A clear evolution from the AFM to the FM state is seen with increasing Co substitution. In Co-free MnNiGe, the AFM M (H) curve with a kink at 13 kOe corresponds to the instability of the spiral AFM structure of stoichiometric MnNiGe.[18,21,29] For $x = 0.1$, an increased slope of the M (*H*) curve in low fields is observed with a high magnetization of 72 emu g$^{-1}$ at 50 kOe. However, the competition between FM and AFM interactions remains strong (also consistent with Fig. 3(b)), which makes it difficult for the material to be saturated. In the sample with *x* = 0.2, the AFM interaction is completely overcome and a typical FM M (*H*) curve, with a low saturation field of about 5 kOe and a saturation magnetization of 70 emu g$^{-1}$, is obtained. This means that only 20% Co substitution for Mn is enough to give rise to a complete FM ground state in MnNiGe martensites. In the martensite of stoichiometric MnNiGe, the magnetic moments are only located on the Mn atoms with a value of 2.8 $\mu_B$.[17] In this alloyed system, the introduced Co atoms directly occupy the substituted Mn sites according to the atomic site-preference rule.[29,30] Assuming that Mn moments remain basically unchanged, a reasonable value of about 0.55 $\mu_B$ for the Co moment can be derived from the sample with *x* = 0.2 (molecular moment = 2.35 $\mu_B$). It is strongly suggested that, over a similar, shortened interatomic distance, the exchange interaction between Co and Mn atoms is prone to be positive and to establish FM coupling, while that between Mn and Mn moments is prone to spiral AFM coupling, as illustrated in insets to Fig. 4 (supplementary Fig. S3).[36] Based on the local Co-4Mn atomic configuration in this Co-substituted MnNiGe, the FM coupling between Co-Mn atoms enables the rapid establishment of ferromagnetism in AFM martensite of MnNiGe.

We present a structural and magnetic phase diagram in Fig. 5 that shows a clear picture of the magnetostructural coupling in MnNiGe-CoNiGe. Compared with the MnNiGe-MnCoGe system (Fig. 1(a)), the temperature gap (blue region) between $T_t$ and $T_N^M$ in the MnNiGe-CoNiGe system is rapidly closed. In addition, $T_t$ of MT decreases below $T_C^A$. Simultaneously, the AFM-FM conversion in the martensite phase is also accomplished. With a span of $T_t$ over the two magnetic-ordering temperatures ($T_N^M$ and $T_C^A$), a TW covering more than 200 K from room temperature to about 120 K is opened. These achievements jointly promote the magnetostructural transition. Within the TW, the MT coupled with a PM-to-FM magnetic transition generates a large $\Delta M$. The expectations proposed in Fig. 1(b) can be effectively realized by the isostructural alloying scheme in the MnNiGe-CoNiGe system.

On the basis of the magnetostructural transition in the obtained TW, the magnetoresponsive effects of the sample with $x$ = 0.1 are presented in Fig. 6. As shown in Fig. 6(a), the metamagnetization behavior (indicated by short arrows) in the isothermal M ($H$) curves is observed between 242 and 236 K in the increasing magnetic field, showing the occurrence of magnetic-field-induced MT from the PM austenite to the FM martensite. With higher magnetization, the martensite gains higher Zeeman energy. This results in a larger decrease in the total free energy and, subsequently, a more stable martensite in the magnetic field.

Shown in Fig. 6(b) is the associated magnetic-entropy changes ($\Delta S_m$) across the MT of $x$ = 0.1. The $\Delta S_m$ values were calculated using the Maxwell relation[33] and the isothermal M ($H$) curves shown in Fig. 6(a). To avoid a physically impossible result for the magnetostructural transition with a thermal hysteresis, the loop process method[34] was adopted in the measurement of the isothermal magnetization. Due to the descent of the magnetic entropy in the PM-to-FM MT, a negative (normal) MCE was obtained. In this sample there are maximal $\Delta S_m$ values of -15 and -40 J kg$^{-1}$ K$^{-1}$, respectively, for field changes ($\Delta H$) of 20 and 50 kOe. These values are much larger than those of many well-known magnetocaloric materials with magnetoelastic or magnetostructural transitions. Primarily, this giant MCE is ascribed to the sharp first-order magnetostructural transition with a large $\Delta M$. In addition, it originates from the characteristics of the phase transitions in the established TW. When the MT occurs at 236 K, as shown in Figs. 3 and 5, the parent phase is far above its intrinsic $T_C^A$ (~129 K) and is thus in a PM state with highly disordered spins, and meanwhile the produced martensite is far below its intrinsic $T_C^M$ (~356 K) and is thus in a FM state with highly ordered spins. A very large change in magnetic entropy therefore takes place in this TW. In sharp contrast, as the MT of $x = 0.2$ goes out of (below) the TW, the $\Delta S_m$ in the FM-to-FM transition remarkably decreases to -5 J kg$^{-1}$ K$^{-1}$ ($\Delta H$ = 50 kOe) (supplementary Fig. S4).[36]

As another important measure of the magnetocaloric materials, the refrigerant capacity (RC) of $x = 0.1$ is evaluated from Fig. 6(b) by using the approach proposed in Ref. 35. The RC is defined as $RC = \int_{T_1}^{T_2} |\Delta S_m| \, dT$, where $T_1$ and $T_2$ are two temperatures of the half maximum of $\Delta S_m (T)$ peak, respectively. A value of about 159 J kg$^{-1}$ is produced, indicating that a large capacity of performing the work in a refrigeration cycle can be expected. Here it should be stated that a magnetic hysteresis behavior was confirmed in the magnetization measurements. The related hysteresis loss will lead to a decrease in the RC.[5] Further efforts should be made to minimize the hysteresis loss of this material in the future.

For many martensitic alloys, the MT occurs from the strong FM parent phase to the relatively weak magnetic (FM, AFM or PM) martensite phase,[4,12,13,15,19] with the heat processes of the structural and magnetic transitions in opposition. The MCEs occurring during these transitions are counteracted by the opposing heat processes, which were previously described.[6,8] Here, for the PM/FM-type MT in the TW (Fig. 3(b)), the enthalpy changes of the combined structural and magnetic transitions always maintain the same sign. We believe that this "locked" caloric behavior could enhance the efficiency of magnetic cooling applications.

In conclusion, we have demonstrated a simple and effective scheme and established a broad Curie-temperature window in MnNiGe-CoNiGe system. Based on the PM/FM-type magnetostructural transition occurring in the window, the giant MCE enables $Mn_{1-x}Co_xNiGe$, an Mn-based material that is free of rare-earth elements, to have a great potential for the solid-state magnetic cooling. The Curie-temperature window provides a way for the large $\Delta M$ to maximize the magnetic-energy change in magnetostructural phase transitions. The isostructural alloying that takes into consideration phase stability and magnetism of the selected isostructures can be properly applied as an active method to simultaneously manipulate the phase transition and the magnetic structure of a host material.

This work is supported by the National Basic Research Program of China (973 Program: 2012CB619405) and the National Natural Science Foundation of China (51021061 and 11174352, 51171206). Dr. Virgina Unkefer improved the English.

[36] See supplementary material at [URL will be inserted by AIP] for the DSC thermal analysis of $T_t$ and $T_N^M$ for stoichiometric MnNiGe. The M (*H*) curves at low temperatures were measured to verify the complex magnetic behavior of

$Mn_{0.9}Co_{0.1}NiGe$. The atomic site preference and related magnetic coupling were further analyzed in Co-substituted MnNiGe martensite. Besides, the MCE of $Mn_{0.8}Co_{0.2}NiGe$ was measured and discussed with a comparison with that of $Mn_{0.9}Co_{0.1}NiGe$.

TABLE I. Phase structures, lattice parameters, martensitic-transition temperatures and magnetic properties of Mn$_{1-x}$Co$_x$NiGe at room temperature.

| $x$ | Structure | $a$ (Å) | $b$ (Å) | $c$ (Å) | $T_C^A$ (K) | $T_N^M$ (K) | $T_t$ (K) | $\Delta M$ (emu g$^{-1}$) |
|---|---|---|---|---|---|---|---|---|
| 0 | *Ortho.* | 6.0275 | 3.7449 | 7.0780 | | 356 | 460 | |
| 0.1 | *Hex.* | 4.0718 | | 5.3598 | | | 236 | 53.0 |
| 0.2 | *Hex.* | 4.0582 | | 5.3413 | 129 | | 93 | 26.0 |
| 0.3 | *Hex.* | 4.0332 | | 5.2947 | 124 | | | |

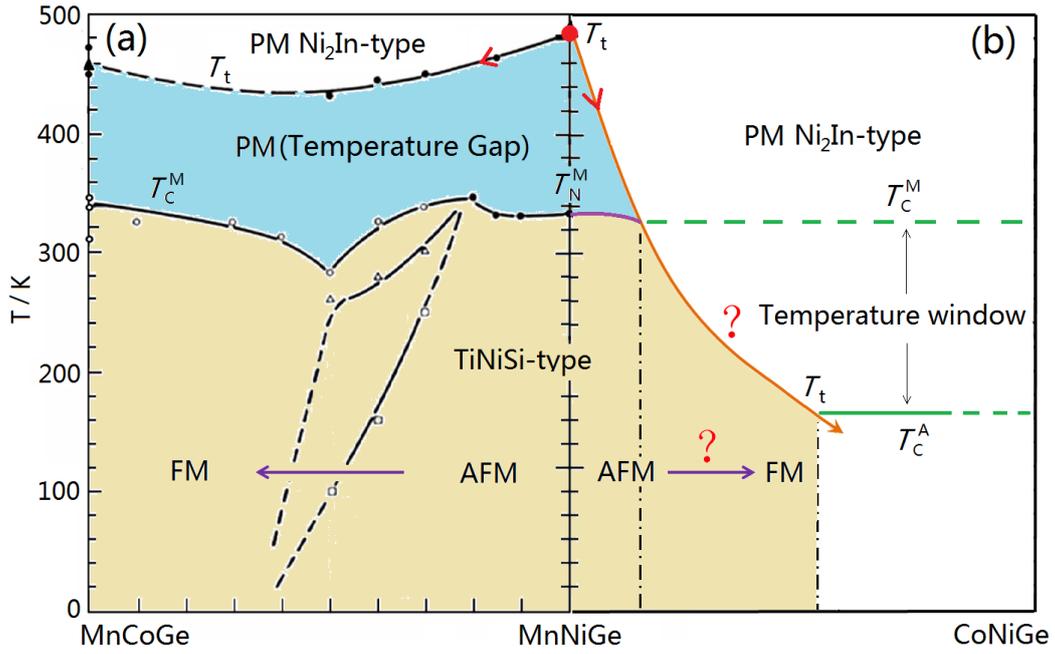

FIG. 1. (a) Structural and magnetic phase diagram of the MnNiGe-MnCoGe system described in ref. 29. Reprinted with permission from Nizioł, *et al.*, J. Magn. Magn. Mater., **27**, 281 (1982). Copyright 1982 Elsevier. (b) Schematic phase diagram of the MnNiGe-CoNiGe system.

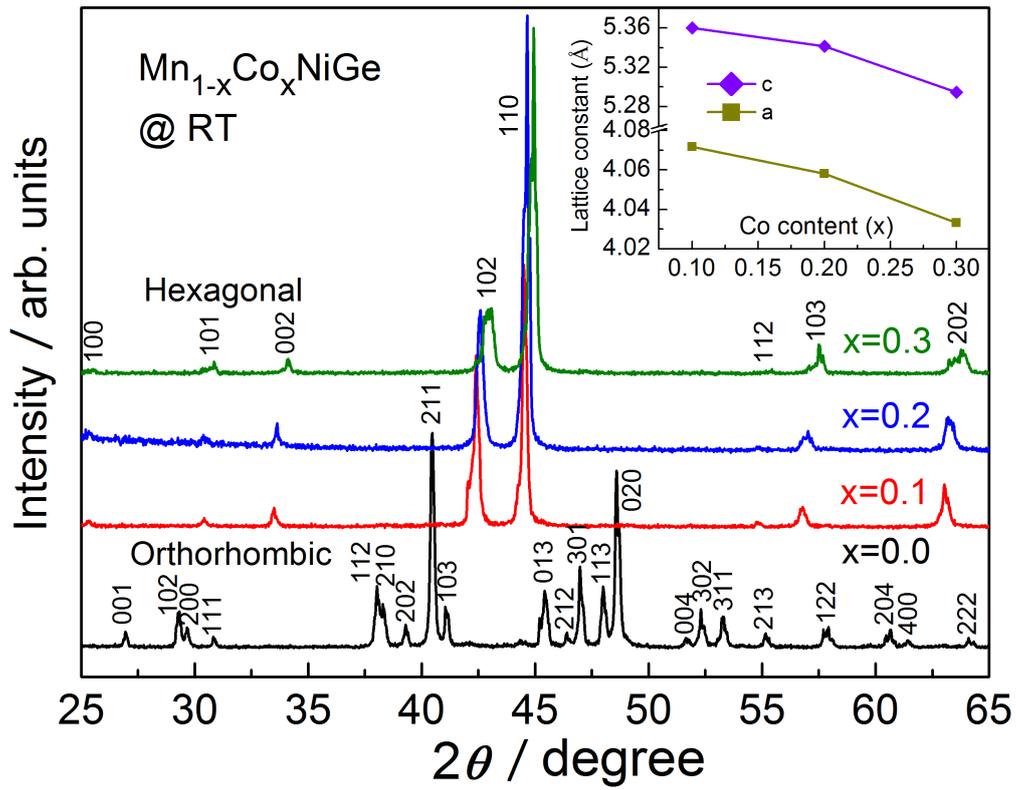

FIG. 2. Room-temperature XRD patterns of Mn$_{1-x}$Co$_x$NiGe. Inset shows the composition-dependent lattice parameters (*a*, *c* axes) of austenite phase.

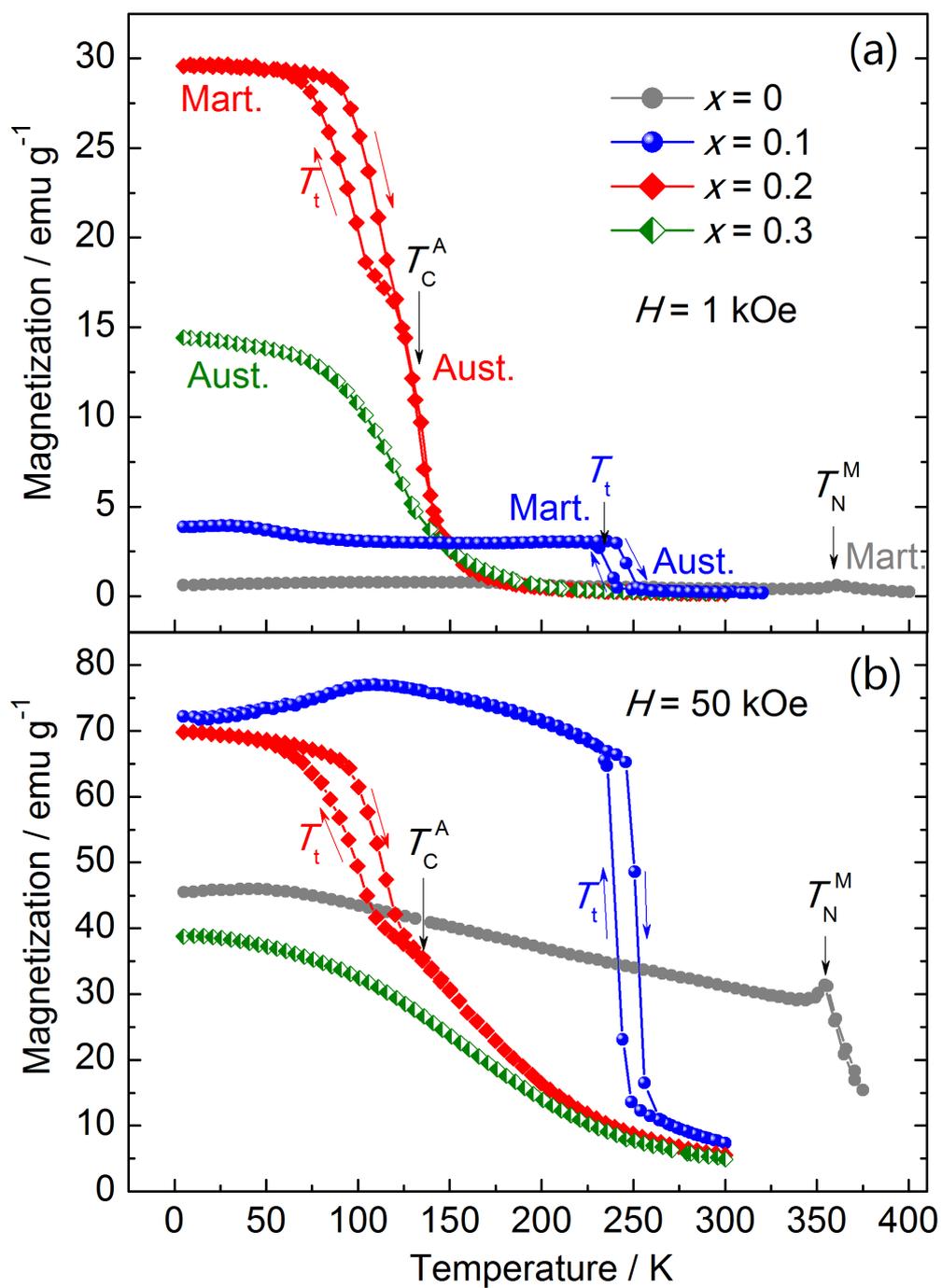

FIG. 3. M (*T*) curves of Mn$_{1-x}$Co$_x$NiGe in low (a) and high (b) magnetic fields.

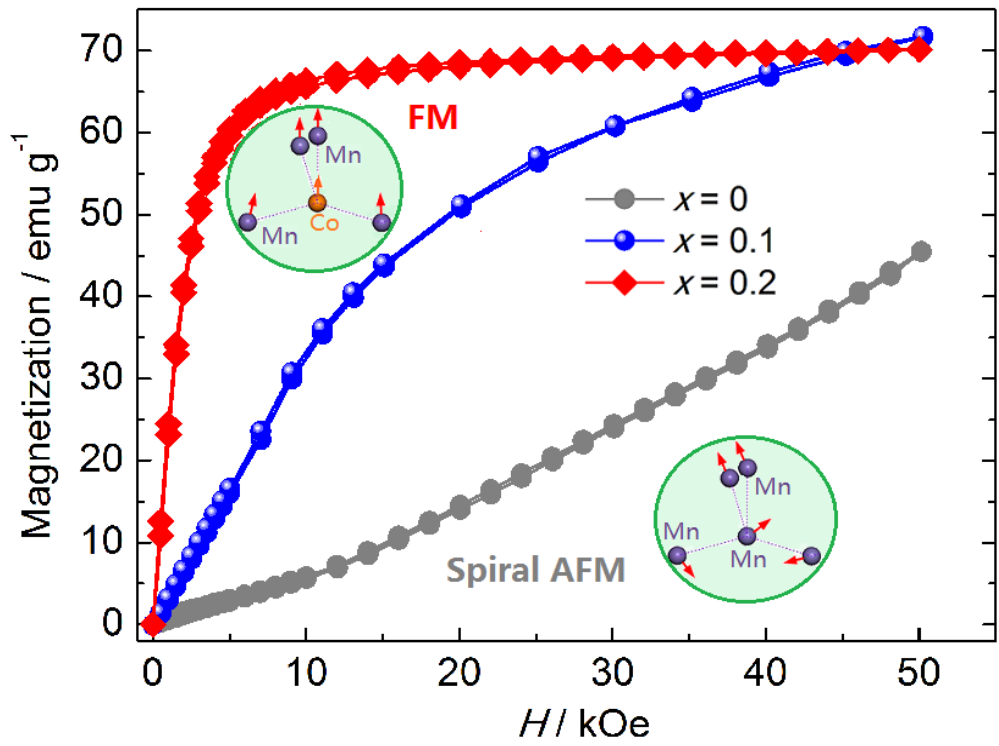

FIG. 4. Isothermal M (*H*) curves of Mn$_{1-x}$Co$_x$NiGe martensite phases at 5 K. The insets illustrate the spiral AFM coupling of Mn-Mn pair and the FM coupling of Co-Mn pair.

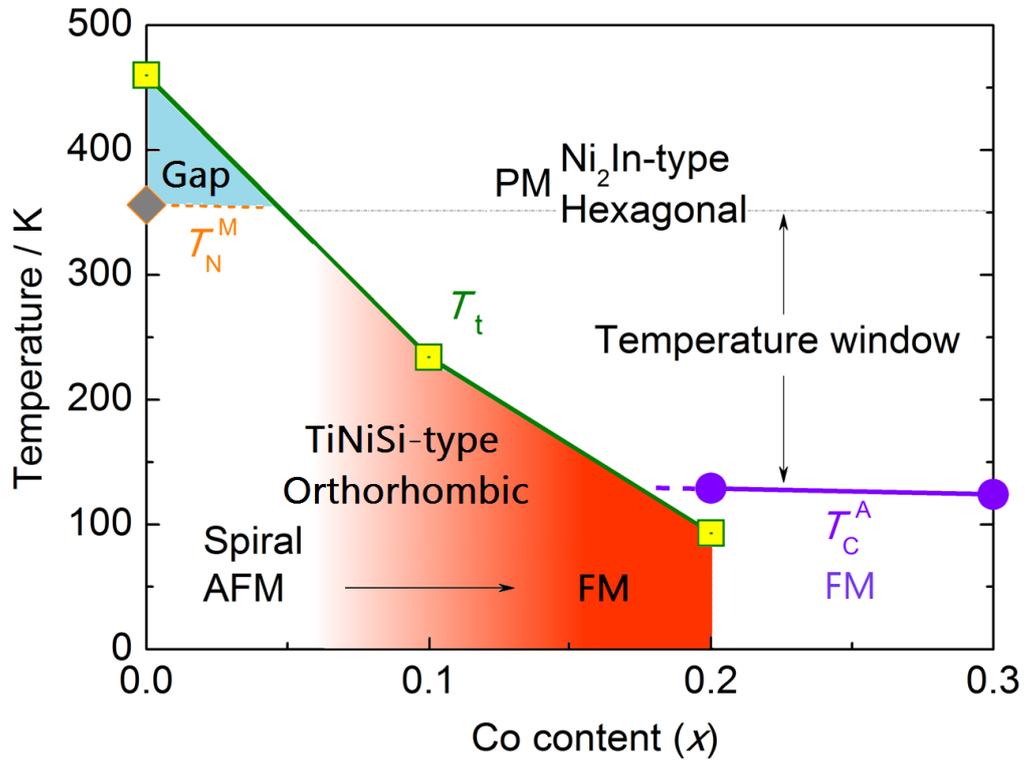

FIG. 5. Magnetostructural phase diagram of Mn$_{1-x}$Co$_x$NiGe. The blue region shows the temperature gap between $T_t$ and $T_N^M$. The gradually increasing intensity of the red color demonstrates the AFM-FM conversion of martensites as a function of Co content.

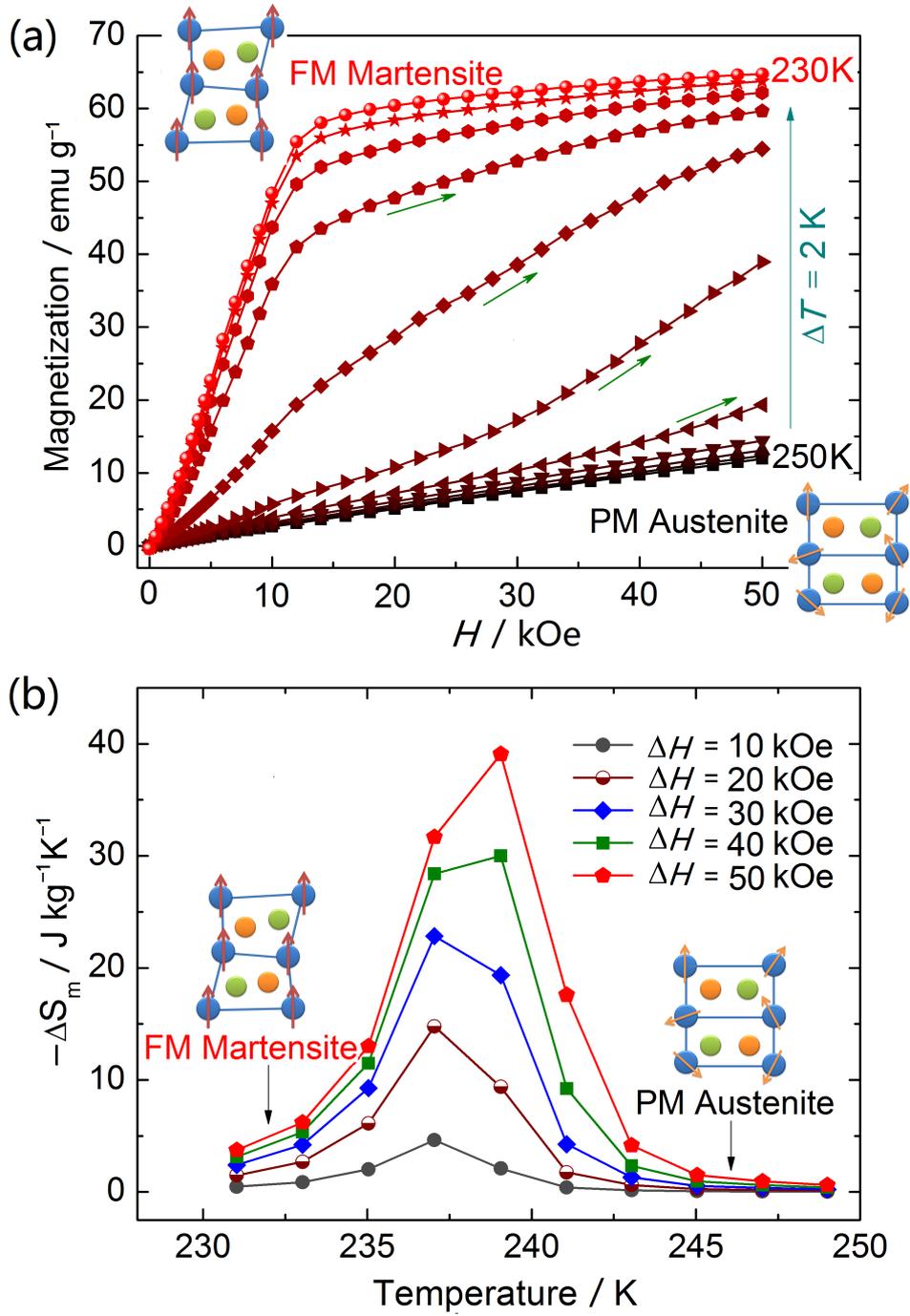

FIG. 6. Magnetoresponsive effects of Mn$_{0.9}$Co$_{0.1}$NiGe. (a) The magnetic-field-induced MT at various temperatures. (b) Isothermal magnetic-entropy changes in various field changes. The insets are schematics of the crystalline and magnetic structures of austenite and martensite.

21